\documentclass[pra,twocolumn, aps, preprintnumbers,superscriptaddress]{revtex4-2}
\usepackage[colorlinks=true,citecolor=blue,urlcolor=black]{hyperref}
\usepackage{verbatim}
\usepackage{amsmath}
\usepackage{latexsym}
\usepackage{revsymb}
\usepackage{yfonts}
\usepackage{ifthen}

\usepackage{natbib}
\usepackage{amsfonts}
\usepackage{amsmath}
\usepackage{amssymb}
\usepackage{amsthm}
\usepackage{graphicx}
\usepackage{bm}
\usepackage{bbm}
\usepackage{epsfig,color,amssymb}
\usepackage{subfigure}
\usepackage{amsfonts}
\usepackage{amscd}
\usepackage{amsmath}
\usepackage{multirow}
\usepackage{chemarrow}
\usepackage{dcolumn}
\usepackage{bm}
\usepackage{graphicx}
\usepackage{enumerate}
\usepackage{epsfig}
\usepackage{subfigure}
\usepackage{xcolor}
\usepackage{multirow}
\usepackage{ulem}
\usepackage{braket}
\usepackage{comment}
\usepackage{enumitem}
\usepackage{amsthm}
\renewcommand{\emph}[1]{\textit{#1}}

\begin{document}
	
	\title{Finite-key Analysis for Quantum Conference Key Agreement with Asymmetric Channels}

	\author{Zhao Li}
	\author{Xiao-Yu Cao}
	\author{Chen-Long Li}
	\author{Chen-Xun Weng}
	\author{Jie Gu}	
	\author{Hua-Lei Yin}\email{hlyin@nju.edu.cn}
	\author{Zeng-Bing Chen}\email{zbchen@nju.edu.cn}
	\affiliation{National Laboratory of Solid State Microstructures, School of Physics and Collaborative Innovation Center of Advanced Microstructures, Nanjing University, Nanjing 210093, China}
	
	
	\begin{abstract}
	 As an essential ingredient of quantum networks, quantum conference key agreement (QCKA) provides unconditional secret keys among multiple parties, which enables only legitimate users to decrypt the encrypted message. Recently, some QCKA protocols employing twin-field was proposed to promote transmission distance. These protocols, however, suffer from relatively low conference key rate and short transmission distance over asymmetric channels, which demands a prompt solution in practice. Here, we consider a tripartite QCKA protocol utilizing the idea of sending-or-not-sending twin-field scheme and propose a high-efficiency QCKA over asymmetric channels by removing the symmetry parameters condition. Besides, we provide a composable finite-key analysis with rigorous security proof against general attacks by exploiting the entropic uncertainty relation for multiparty system. Our protocol greatly improves the feasibility to establish conference keys over asymmetric channels.
	\end{abstract}

	\maketitle
	\section{INTRODUCTION}
	Remote distribution of secret key is an essential task of quantum cryptographic network. Quantum key distribution (QKD) \cite{bennett1984proceedings, Ekert1991Quantum, lo1999unconditional, shor2000simple, RevModPhys.92.025002} which allows two remote users to share unconditionally secure key has achieved relatively mature development \cite{wang2005beating,lo2005decoy,lo2012measurement,yin2014long,zhou2016making,yin2016measurement,lucamarini2018overcoming, wang2018twin,ma2018phase, lin2018simple, yin2019measurement, cui2019twin, curty2019simple, yin2019coherent, xu2020sending}. To extend QKD to multiparty scenarios, one intuitive way is to refer to classical conference key agreement \cite{diffie1976new, burmester1994secure}. Therefore, quantum conference key agreement (QCKA) \cite{bose1998multiparticle, cabello2000multiparty, chen2005conference, matsumoto2007multiparty} has been proposed and become a significant pillar of quantum cryptography. QCKA can offer information-theoretic secure conference key which promises group encryption and decryption for all legitimate users. Specifically, the conference key can be naturally established with distribution of Greenberger-Horne-Zeilinger (GHZ) entangled states \cite{greenberger1989going, mermin1990extreme, bose1998multiparticle} and local projection measurement, which has been proved to precede simple bipartite QKD links \cite{epping2017multi}. Basic laws of quantum mechanics combined with one-time pad encryption \cite{shanno1949communication} guarantee the security of multiparty quantum communication \cite{cabello2000multiparty, Fu2015Long}. Currently, QCKA has been generalized to other variants both in theory
	\cite{chen2016biased, ribeiro2018fully, holz2020genuine, grasselli2018finite, wu2016continuous, ottaviani2019modular, cao2021coherent, Cao2020Open, Murta2020Quantum} and experiment \cite{proietti2021experimental}.

	Although some progresses have been made, the conference key rates of most QCKA protocols are lower than the capacity of quantum channels \cite{das2019universal, pirandola2020general}. To achieve higher conference key rate and longer transmission distance, some protocols inspired by the idea of twin-field QKD were proposed \cite{grasselli2019conference, zhao2020phase, hua2020breaking}. However, all of these protocols neglect the case of asymmetric channels, which makes them impractical. Our paper focuses on the tripartite QCKA protocol \cite{hua2020breaking} employing the sending-or-not-sending scheme \cite{wang2018twin}. The precondition for security requires two senders having the same source parameters (e.g. the intensity and the probability of sending varied coherent pulses). Asymmetric channels with symmetric source parameters will introduce additional intrinsic errors, which inevitably reduce the conference key rate and the transmission distance. In practical, such as intercity communication, the transmission losses of two channels are usually asymmetric.
	
	Inspired by the variants of twin-field QKD over asymmetric channels \cite{yin2019coherent,hu2019sending, zhou2019asymmetric, grasselli2019asymmetric, wang2020simple, zhong2021proof}, we design a high-efficiency QCKA protocol for asymmetric channels. With the similar devices as twin-field QKD \cite{chen2020sending, fang2020implementation}, our protocol has a longer transmission distance, a higher key rate and is more practical than other QCKA protocols \cite{Murta2020Quantum}. We provide a security proof for this protocol by utilizing the entanglement distillation \cite{Fu2015Long} of virtual GHZ-class state. In addition, with the extension of twin-field QKD to QCKA, we perform a finite-key analysis \cite{tomamichel2012tight, lim2014concise, grasselli2018finite, yin2020tight} with multiparty composable security to make our work more practical. 
	
\section{Protocol description}
\begin{figure}[t]
	\centering
	
		\includegraphics[width=8.6cm]{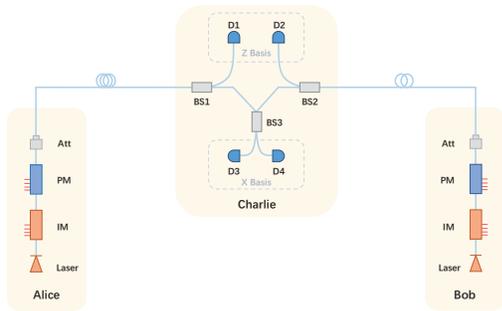}\label{A}

	\caption{The schematic of the asymmetric QCKA protocol. There are two senders, Alice and Bob, including some primary devices as follows. Continuous-wave laser: generation of the global phase stabilized coherent light; intensity modulator (IM): pulse preparation; phase modulator (PM): phase randomization; attenuator (Att): weak-light modulation. They send the pulses to Charlie over their own channels. Charlie, as a receiver, possesses the following devices. Beam splitters BS1 and BS2: passive-basis choice; beam splitter BS3: interference between two pulses; detector D1 and D2: the Z basis measurement; detector D3 and D4: the X basis measurement. }
	\label{LETTER}
\end{figure}

As shown in figure \ref{LETTER}, our asymmetric QCKA protocol includes three legitimate users: Alice, Bob and Charlie with the same devices as the original scheme \cite{hua2020breaking}. We utilize the decoy-state method with three intensities for both senders. We present the detailed protocol as follows. 

\noindent\textbf{Preparation.}—
For each time window, Alice and Bob independently and randomly choose one basis from $\{\mathrm{Z},\ \mathrm{X}\}$. If Alice (Bob) selects Z basis, she (he) encodes the logic bit value $1\ (0)$ with phase-randomized weak coherent pulses $\ket{e^{i\omega_a}\sqrt{\mu_a}}\ (\ket{e^{i\omega_a}\sqrt{\mu_b}})$, whose probability is $t_a\ (t_b)$. She (He) encodes the logic bit value $0\ (1)$ with sending nothing, whose probability is $1-t_a\ (1-t_b)$. If Alice (Bob) selects X basis, she (he) generates phase-randomized weak coherent pulses $\ket{e^{i\theta_a}\sqrt{k_a}}\ (\ket{e^{i\theta_b}\sqrt{k_b}})$, with random global phase $\theta_a,\ \theta_b\in\left[0,\ 2\pi\right)$ and intensity $k_a\in \{\mu_a,\ \nu_a,\ 0\},\ k_b\in \{\mu_b,\ \nu_b,\ 0\}$. They send the pulses to Charlie through two insecure quantum channels that may have different losses. Besides, a constraint for source parameters is required \cite{hu2019sending}, 
\begin{equation}\label{mathematical constraint}
	\frac{\nu_{a}}{\nu_{b}}=\frac{t_a\left(1-t_b\right) \mu_a e^{-\mu_a}}{t_b\left(1-t_a\right) \mu_b e^{-\mu_b}},
\end{equation}
which will be given by security analysis. 

\noindent\textbf{Measurement.}—
For each round, Charlie passively selects a basis from Z and X bases by two beam splitters and performs measurement. For Z basis, Charlie gets logic bit 1 when detector D1 clicks only, and logic bit 0 when detector D2 clicks only. The bit value is 0 when D3 clicks only in X basis while the bit value is 1 when D4 clicks only. 

\noindent\textbf{Reconciliation.}—
Alice, Bob, and Charlie announce the basis information for each successful measurement event, which means that only one detector clicks in each round of measurement. Alice and Bob publicly announce the intensity information unless Alice, Bob, and Charlie all select Z basis. In the case where Alice and Bob choose intensities $\nu_a$, $\nu_b$ respectively and only D3 or D4 clicks, Alice and Bob disclose the global phase. Then they perform post-selected phase-matching, which screens out the effective events satisfying 
$-\delta+r\pi \leq \theta_a-\theta_b-\varphi_{ab} \leq \delta+r\pi \ (\mathrm{mod}~2\pi)$, where $\delta$ is a selected value for a small phase range and $r \in \{0,\ 1\}$. Here $\varphi_{ab}$ is the phase difference between the reference frames of Alice and Bob, which can be obtained by phase post-compensation methods. All classical information is transmitted via the authenticated classical channels.

\noindent\textbf{Parameter estimation.}—
Alice, Bob, and Charlie form raw keys from the data when they all selected Z basis. Utilizing the decoy-state method \cite{wang2005beating, lo2005decoy}, the other cases are exploited to estimate parameters and the information leakage.

\noindent\textbf{Postprocessing.}—
They implement classical error correction, error verification and privacy amplification to distill the final conference key.

\section{Security analysis}

In our protocol, Alice randomly chooses a basis for each round. For Z basis, she prepares $\ket{0}$ and $\ket{e^{i\omega_a}\mu_a}$ by using the vacuum and phase-randomized coherent state with probabilities $1-t_a$ and $t_a$, where $\omega_a\in\left[0,\ 2\pi\right)$ is the randomized phase. For X basis, she randomly generates phase-randomized coherent state $\ket{e^{i\theta_a}\sqrt{\nu_a}}$. Likewise, Bob does the same operations as Alice.

Alice and Bob only reserve joint single-photon states in Z basis, i.e. $\ket{10}_{ab}$ and $\ket{01}_{ab}$, to extract the conference key. The density matrix is 
\begin{equation}
\begin{aligned}
	\rho_{z}^1=&\left[t_{a}\left(1-t_{b}\right) \mu_{a} e^{-\mu_{a}}\ket{10}_{ab}\bra{10}\right.\\
	&\left.+t_{b}\left(1-t_{a}\right) \mu_{b} e^{-\mu_{b}}\ket{01}_{ab} \bra{01}\right]/C,
\end{aligned}
\end{equation}
where $C=t_{a}\left(1-t_{b}\right) \mu_{a} e^{-\mu_{a}}+t_{b}\left(1-t_{a}\right) \mu_{b} e^{-\mu_{b}}$ is the normalization factor.
They implement post-selected phase-matching and only consider the case of joint one-photon states in X basis.
The density matrix is
\begin{equation}
	\rho_x^1=\frac{1}{\nu_{a}+\nu_{b}}(\nu_a\ket{10}_{ab}\bra{10}+\nu_b\ket{01}_{ab}\bra{01}),
\end{equation}
which needs to meet $\rho_x^1=\rho_{z}^1=\rho^1$, so we get equation (\ref{mathematical constraint}). Note that we introduce the phase-matching here to obtain low phase error rate. The decoy state method can guarantee the feasibility of retaining joint single-photon states. Our security analysis only considers perfect sources which can prepare pulses with precise intensity for meeting the constraint of equation (\ref{mathematical constraint}). However, due to the source imperfection in practice, the intensity fluctuations \cite{Wang2009Decoy} will affect Alice’s and Bob’s state preparation, which is still an issue for consideration. The security analysis for more practical cases should be further considered.

Here, we consider a entanglement-swapping virtual protocol \cite{yin2019measurement}, which has a same density matrix as the one mentioned above.  We provide an security proof for our protocol by exploiting the entanglement distillation argument of GHZ-class entangled state \cite{Fu2015Long}. 

\subsection*{Virtual protocol.}

($i$) Alice prepares entangled state 
\begin{equation}
	\begin{aligned}
		\ket{\Phi}_{Aa}&=\sqrt{1-t_{a}}\ket{0}_{A}\ket{0}_{a}+\sqrt{t_{a}\mu_{a}e^{-\mu_{a}}}\ket{1}_{A}\ket{1}_{a},
	\end{aligned}
\end{equation}
where $A$ is the virtual qubit system and $a$ is the optical mode. $\ket{0}$ and $\ket{1}$ are the eigenstates of Z basis while $\ket{+}$ and $\ket{-}$ are the eigenstates of X basis for both virtual qubit system and optical mode, where $\ket{\pm}=\frac{1}{\sqrt2}(\ket{0}\pm\ket{1})$. Bob has the same entangled state.
Alice and Bob send the optical modes $a$ and $b$ to Charlie respectively. 

Here, we introduce an additional virtual step: Alice and Bob jointly perform quantum non-demolition measurement on the state of optical modes $a$ and $b$ to implement photon-number-resolving before they send the state to Charlie. Alice and Bob only reserve the cases of joint single-photon states of modes $a$ and $b$. The joint quantum state between Alice and Bob after this step is 
\begin{equation}
	\begin{aligned}
		\ket{\psi}_{ABab}&=\frac{\sqrt{\nu_{a}}\ket{1010}_{ABab}+\sqrt{\nu_{b}}\ket{0101}_{ABab}}{\sqrt{\nu_{a}+\nu_{b}}} ,
	\end{aligned}
\end{equation}
where $\nu_{a}$ and $\nu_{b}$ meet the constraint of equation (\ref{mathematical constraint}). 
Note that Charlie can use optical modes $\{\ket{01}_{ab},\ \ket{10}_{ab}\}$ to represent the virtual qubit system $\{\ket{0}_C,\ \ket{1}_C\}$. Therefore, we can rewrite the joint quantum state 
\begin{equation}
	\begin{aligned}
		\ket{\phi}_{ABC}=\frac{1}{\sqrt{\nu_{a}+\nu_{b}}}&\left(\sqrt{\nu_a}\ket{1}_A\ket{0}_B\ket{1}_C\right.\\	&\left.+\sqrt{\nu_b}\ket{0}_A\ket{1}_B\ket{0}_C\right),
	\end{aligned}
\end{equation}
which is the GHZ-class entangled states.

($\romannumeral2$) For each round, Alice (Bob) independently and randomly chooses a basis from $\{\mathrm{Z},\ \mathrm{X}\}$ corresponding to the states $\{\ket{0}_{A(B)},\ \ket{1}_{A(B)}\}$ and $\{\ket{+}_{A(B)},\ \ket{-}_{A(B)}\}$. They then perform local projective measurements on their kept virtual qubits. Charlie measures the qubit system $C$ in the Z or X basis to obtain his logic bit. 
($\romannumeral3$) Alice, Bob, and Charlie announce the basis information for each round of measurement. 
($\romannumeral4$) Alice, Bob, and Charlie extract raw keys from the data of Z basis while the data of X basis for estimation of information leakage. 
($\romannumeral5$) They distill final key by exploiting the classical postprocessing.

For the above virtual protocol, the joint single-photon state of modes $a$ and $b$ has the same density matrix as $\rho^1$ when Alice and Bob both measures the kept qubit systems in the $Z$ or $X$ basis. Therefore, for the single-photon state, one cannot distinguish the practical protocol as shown in figure~\ref{LETTER} and the virtual protocol.
We let the probability of Alice and Bob choosing Z basis $p_{z_a},\ p_{z_b}\approx1$ in the asymptotic limit. The secret key rate of our protocol is
\begin{equation}\label{m1}
	\begin{aligned}
		R=&\left[t_a(1-t_b)e^{-\mu_a}+t_b(1-t_a)e^{-\mu_b}\right]Y_{0}^{z}\\
		&+\left[t_a(1-t_b)\mu_ae^{-\mu_a}+t_b(1-t_a)\mu_b e^{-\mu_b}\right]\\ 
		&\times Y_{1}^{z}\left[1-h\left(e_{1}^{\mathrm{ph}}\right)\right]-\xi_{\mathrm{EC}},
	\end{aligned}
\end{equation}
where $Y_{0}^{z}$ and $Y_{1}^{z}$ are the yields when Alice and Bob both send the vacuum state and joint single-photon states given that all users choose Z basis respectively. $e_{1}^{\mathrm{ph}}$ is the phase error rate with all users choosing Z basis, which can be estimated by the bit error rate of X basis. $\xi_{\mathrm{EC}}=Q^zfh(E^z)$ is the information leakage in classical error correction, where $Q^z$ is the gain of they all selecting Z basis, $f$ is the efficiency of error correction, and $E^z$ is the bit error rate of Z basis. $h(x)=-x\log{(x)}-(1-x)\log{(1-x)}$ is the binary Shannon entropy function. 

The yields $Y_{0}^{d}=Q_{00}^{d}$ and $Y_{1}^{d}$ can be estimated by 
\begin{equation}
	Y_{1}^{d} \geq \frac{\nu_a}{\nu_a+\nu_b}Y_{10}^d+\frac{\nu_b}{\nu_a+\nu_b}Y_{01}^d,
\end{equation}
where $Y_{10}^{d}$ is the yield that Alice sends single-photon state while Bob sends the vacuum state, and $Y_{01}^{d}$ is the yield that Alice sends the vacuum state while Bob sends the single-photon state. $d$ denotes all users choose D basis, with $\mathrm{D}\in\{\mathrm{Z},\ \mathrm{X}\}$. 
The yields $Y_{10}^{d}$ and $Y_{01}^{d}$ can be bounded by exploiting the decoy-state method \cite{wang2005beating, lo2005decoy},
\begin{equation}
	Y_{10}^{d} \geq \frac{\mu_a}{\mu_a \nu_a-\nu_a^{2}}\left[e^{\nu_a} Q_{\nu_a0}^{d}-\frac{\nu_a^{2}}{\mu_a^{2}} e^{\mu_a} Q_{\mu_a0}^{d}-\frac{\mu_a^{2}-\nu_a^{2}}{\mu_a^{2}} Q_{00}^{d}\right],
\end{equation}
\begin{equation}
	Y_{01}^{d} \geq \frac{\mu_b}{\mu_b \nu_b-\nu_b^{2}}\left[e^{\nu_b} Q_{0\nu_b}^{d}-\frac{\nu_b^{2}}{\mu_b^{2}} e^{\mu_b} Q_{0\mu_b}^{d}-\frac{\mu_b^{2}-\nu_b^{2}}{\mu_b^{2}} Q_{00}^{d}\right],
\end{equation}
where $Q_{k_ak_b}^{d}$ is the gain of Charlie's D basis given that Alice and Bob send pulses with intensities $k_a$ and $k_b$ respectively. The phase error rate can be bounded by
\begin{equation}
	e_{1}^{\mathrm{ph}} \leq \frac{1}{(\nu_a+\nu_b)Y_{1}^{x}}\left(e^{\nu_a+\nu_b}E_{\nu_a\nu_b}^{\mathrm{pm}}Q_{\nu_a \nu_b}^{\mathrm{pm}}-\frac{1}{2} Y_{0}^{x}\right),
\end{equation}
where $E_{\nu_a\nu_b}^{\mathrm{pm}}$ and $Q_{\nu_a \nu_b}^{\mathrm{pm}}$ are the bit error rate and gain of Charlie selecting X basis when Alice chooses intensity $\nu_a$ while Bob chooses intensity $\nu_b$ and they successfully perform the postselected phase-matching.

\section{Finite-key analysis}
Here, we give the key rate formula of our asymmetric QCKA protocol considering the effect of the finite key size \cite{tomamichel2012tight, lim2014concise,yin2020tight}. The protocol is $\varepsilon_{\rm sec}$-secret if the secret key of length $l$ satisfies \cite{tomamichel2011uncertainty,  Fu2015Long, grasselli2018finite}
\begin{equation}\label{keylength}
	l=\underline{s}_0^{z}+\underline{s}_1^{z}[1-h(\overline{e}_{1}^{\mathrm{ph}})]-\lambda_{\rm EC} -\log_2\frac{4}{\varepsilon_{\rm cor}}
	-6\log_2\frac{26}{\varepsilon_{\rm sec}},
\end{equation}
where $s_0^{z}$ and $s_1^{z}$ are the numbers of successful events that Alice and Bob both send the vacuum state and joint single-photon states when all users choose Z basis respectively. Note that we define $x$ as the observed value, $x^*$ as the expected value, $\underline{x}$ and $\overline{x}$ as the lower and upper bound of $x$. Here, the protocol is $\varepsilon_{\rm cor}$-correct \cite{grasselli2018finite}, where $\varepsilon_{\rm cor}$ is the probability that the error correction fails. $\lambda_{\rm EC}=n^z f h\left(E^z\right)$ is the error correction leakage, with the number of event $n^z$ that all users choose Z basis.

We can use the announced event to estimate our unknown parameters in equation (\ref{keylength}). When Alice and Bob sent pulses with intensities $k_a$ and $k_b$, we let $N_{k_ak_b}$ be the number of the announced event, e.g. 
$\begin{aligned}
    N_{00}=&\left[\left(1-p_{z_a}\right)\left(1-p_{z_b}\right)p_{0_a}p_{0_b}+\left(1-p_{z_a}\right)p_{z_b}p_{0_a}(1-t_b)\right.\\
    &\left.+p_{z_a}\left(1-p_{z_b}\right)(1-t_a)p_{0_b}\right]N
\end{aligned}$, where $p_{0_a}$ and $p_{0_b}$ are the probabilities of choosing the vacuum state given that Alice and Bob select X basis. $n_{k_ak_b}^{d}$ is the observed number of effective event measured by basis $\mathrm{D}\in\{\mathrm{X},\ \mathrm{Z}\}$. By exploiting the decoy-state method for finite sample sizes, we can derive the lower bound on the expected numbers of vacuum events 
\begin{equation}
\begin{aligned}
    	s_{0}^{z*} \geq \underline{s}_{0}^{z*}=&p_{z_a}p_{z_b}\left[t_a(1-t_b)e^{-\mu_a}+t_b(1-t_a)e^{-\mu_b}\right]\\
	&\times N\frac{\underline{n}_{00}^{z*}}{N_{00}},
\end{aligned}
\end{equation}
and joint single-photon events 
\begin{equation}
	s_{1}^{z*} \geq \underline{s}_{1}^{z*}=\underline{s}_{10}^{z*}+ \underline{s}_{01}^{z*},
\end{equation}
where $N$ is the total number of the all events. Here, the lower bound on expected numbers of two kinds of joint single-photon events, $\underline{s}_{10}^{z*}$ and $\underline{s}_{01}^{z*}$, can be given by 
\begin{equation}
	\begin{aligned}
		s_{10}^{z*} \geq \underline{s}_{10}^{z*}=&\frac{t_a(1-t_b)\mu_a^2 e^{-\mu_a}p_{z_a}p_{z_b}N}{\mu_a \nu_a-\nu_a^{2}}\\
		&\times\left[e^{\nu_a} \frac{\underline{n}_{\nu_a0}^{z*}}{N_{\nu_a0}}-\frac{\nu_a^{2}}{\mu_a^{2}} e^{\mu_a} \frac{\overline{n}_{\mu_a0}^{z*}}{N_{\mu_a0}}-\frac{\mu_a^{2}-\nu_a^{2}}{\mu_a^{2}} \frac{\overline{n}_{00}^{z*}}{N_{00}}\right],
	\end{aligned}
\end{equation}
\begin{equation}
	\begin{aligned}
		s_{01}^{z*} \geq \underline{s}_{01}^{z*}=&\frac{t_b(1-t_a)\mu_b^2 e^{-\mu_b}p_{z_a}p_{z_b}N}{\mu_b \nu_b-\nu_b^{2}}\\
		&\times\left[e^{\nu_b} \frac{\underline{n}_{0\nu_b}^{z*}}{N_{0\nu_b}}-\frac{\nu_b^{2}}{\mu_b^{2}} e^{\mu_b} \frac{\overline{n}_{0\mu_b}^{z*}}{N_{0\mu_b}}-\frac{\mu_b^{2}-\nu_b^{2}}{\mu_b^{2}} \frac{\overline{n}_{00}^{z*}}{N_{00}}\right].
	\end{aligned}
\end{equation}

The successful probability of post-selected phase-matching is $p_{\mathrm{pm}}=\frac{2\delta}{\pi}$. We can also work out the lower bound on the expected number of joint single-photon events after successful post-selected phase-matching $\underline{s}_{1}^{\mathrm{pm}*}$, 
\begin{equation}
	s_{1}^{\mathrm{pm}*} \geq \underline{s}_{1}^{\mathrm{pm}*}=\underline{s}_{10}^{\mathrm{pm}*}+ \underline{s}_{01}^{\mathrm{pm}*}.
\end{equation}
Similarly, the lower bound on expected numbers of two kinds of joint single-photon events, $\underline{s}_{10}^{\mathrm{pm}*}$ and $\underline{s}_{01}^{\mathrm{pm}*}$, can be given by
\begin{equation}
	\begin{aligned}
		s_{10}^{\mathrm{pm}*}\geq\underline{s}_{10}^{\mathrm{pm}*}=
		&\frac{(1-p_{z_a})(1-p_{z_b})p_{\nu_a}p_{\nu_b}p_{\mathrm{pm}}\mu_a\nu_a e^{-(\nu_a+\nu_b)}N}{\mu_a \nu_a-\nu_a^{2}}\\
		&\times\left[e^{\nu_a} \frac{\underline{n}_{\nu_a0}^{x*}}{N_{\nu_a0}}-\frac{\nu_a^{2}}{\mu_a^{2}} e^{\mu_a} \frac{\overline{n}_{\mu_a0}^{x*}}{N_{\mu_a0}}-\frac{\mu_a^{2}-\nu_a^{2}}{\mu_a^{2}} \frac{\overline{n}_{00}^{x*}}{N_{00}}\right],
	\end{aligned}
\end{equation}
\begin{equation}
	\begin{aligned}
		s_{01}^{\mathrm{pm}*}\geq\underline{s}_{01}^{\mathrm{pm}*}=
		&\frac{(1-p_{z_a})(1-p_{z_b})p_{\nu_a}p_{\nu_b}p_{\mathrm{pm}}\mu_b\nu_b e^{-(\nu_a+\nu_b)}N}{\mu_b \nu_b-\nu_b^{2}}\\
		&\times\left[e^{\nu_b} \frac{\underline{n}_{0\nu_b}^{x*}}{N_{0\nu_b}}-\frac{\nu_b^{2}}{\mu_b^{2}} e^{\mu_b} \frac{\overline{n}_{0\mu_b}^{x*}}{N_{0\mu_b}}-\frac{\mu_b^{2}-\nu_b^{2}}{\mu_b^{2}} \frac{\overline{n}_{00}^{x*}}{N_{00}}\right],
	\end{aligned}
\end{equation}
Where we denote the probability of Alice and Bob choosing intensities $\nu_a$ and $\nu_b$ given that they all select X basis as $p_{\nu_a}$ and $p_{\nu_b}$. The upper and lower bound of all the above expected values can be obtained by using the variant of Chernoff bound with the announced observed values \cite{yin2019finite} for each parameter with failure probability $\varepsilon_{\rm sec}/26$. Once acquiring the lower bound of expected values $\underline{s}_{0}^{z*}$, $\underline{s}_{1}^{z*}$ and $\underline{s}_{1}^{\mathrm{pm}*}$, one can exploit the Chernoff bound \cite{yin2019finite} to calculate the lower bound of corresponding observed values $\underline{s}_{0}^{z}$, $\underline{s}_{1}^{z}$ and $\underline{s}_{1}^{\mathrm{pm}}$ for each parameter with failure probability $\varepsilon_{\rm sec}/26$. 

By utilizing the decoy-state method, we can also estimate the upper bound of the amount of bit error $\overline{t}_{1}^{\mathrm{pm}}$ associated with the joint single-photon events after successful post-selected phase-matching \cite{yin2020experimental}, 
\begin{equation}
	t_{1}^{\mathrm{pm}} \leq\overline{t}_{1}^{\mathrm{pm}}=m_{\nu_a\nu_b}^{\mathrm{pm}}-\underline{t}_{0}^{\mathrm{pm}},
\end{equation}
with
\begin{equation}
	\underline{t}_{0}^{\mathrm{pm}*}=(1-p_{z_a})(1-p_{z_b})p_{\nu_a}p_{\nu_b}p_{\mathrm{pm}}e^{-(\nu_a+\nu_b) }N\frac{\underline{n}_{00}^{x*}}{2N_{00}},
\end{equation}
where $m_{\nu_a\nu_b}^{\mathrm{pm}}$ is the observed number of bit error after the post-selected phase-matching. Note that expected value of bit errors on vacuum events satisfies $m_{00}^{x*}=n_{00}^{x*}/2$. The lower bound on the observed number of bit error in the vacuum state $\underline{t}_{0}^{\mathrm{pm}}$ can also be obtained by exploiting the Chernoff bound.  
The hypothetically observed phase error rate associated with the joint single-photon events in Z basis can be obtained by using the random sampling without replacement \cite{yin2019finite}, 
\begin{equation}
	\overline{e}_{1}^{\mathrm{ph}}= \frac{\overline{t}_1^{\mathrm{pm}}}{\underline{s}_1^{\mathrm{pm}}}+\gamma\left(\underline{s}_1^{z},\underline{s}_1^{\mathrm{pm}},\frac{\overline{t}_1^{\mathrm{pm}}}{\underline{s}_1^{\mathrm{pm}}},\varepsilon_{\rm sec}/26\right),
\end{equation}
where $\gamma$ is the parameter of the random sampling without replacement with failure probability $\varepsilon_{\rm sec}/26$. The variant of Chernoff bound is used twelve times, the Chernoff bound is used four times and the random sampling without replacement is used one time. Setting each error term to a common value, we get the factor is 26, including 9 error terms due to the smooth min-entropy estimation \cite{lim2014concise}.

\section{Simulation details}
\begin{figure}[t!]
	\centering
		\includegraphics[width=8.6cm]{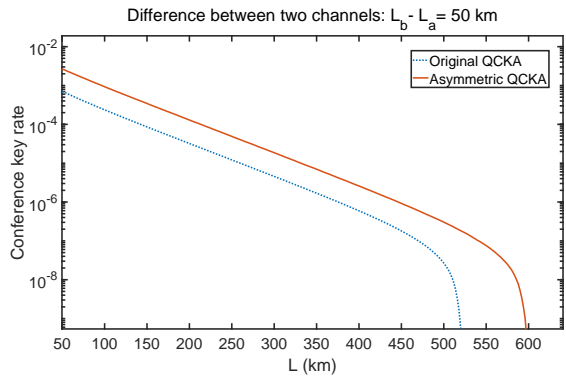}

	\caption{Comparison of the secure conference key rate between our protocol and the original protocol in infinite-key asymptotic limit. $L=L_a+L_b$ is the total transmission distance. Here, the difference in length between two channels is fixed at 50 km, i.e. $L_b-L_a=50\mathrm{km}$. The typical experimental parameters are listed in table \ref{parameters}.
	}
	\label{wxmc}
\end{figure}

\begin{figure}[t!]
	\centering
		\includegraphics[width=8.6cm]{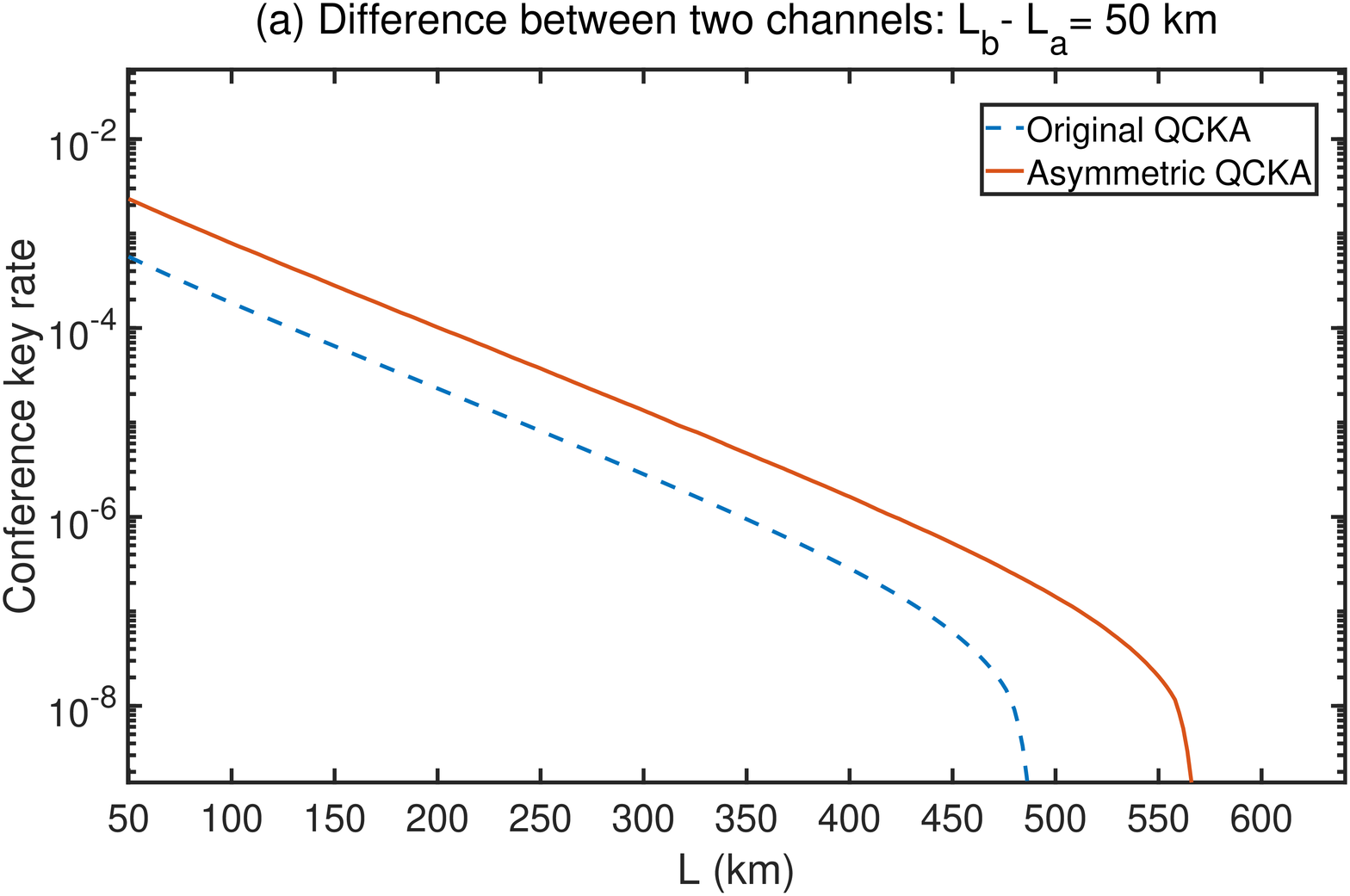}\label{A}

	\quad

		\includegraphics[width=8.6cm]{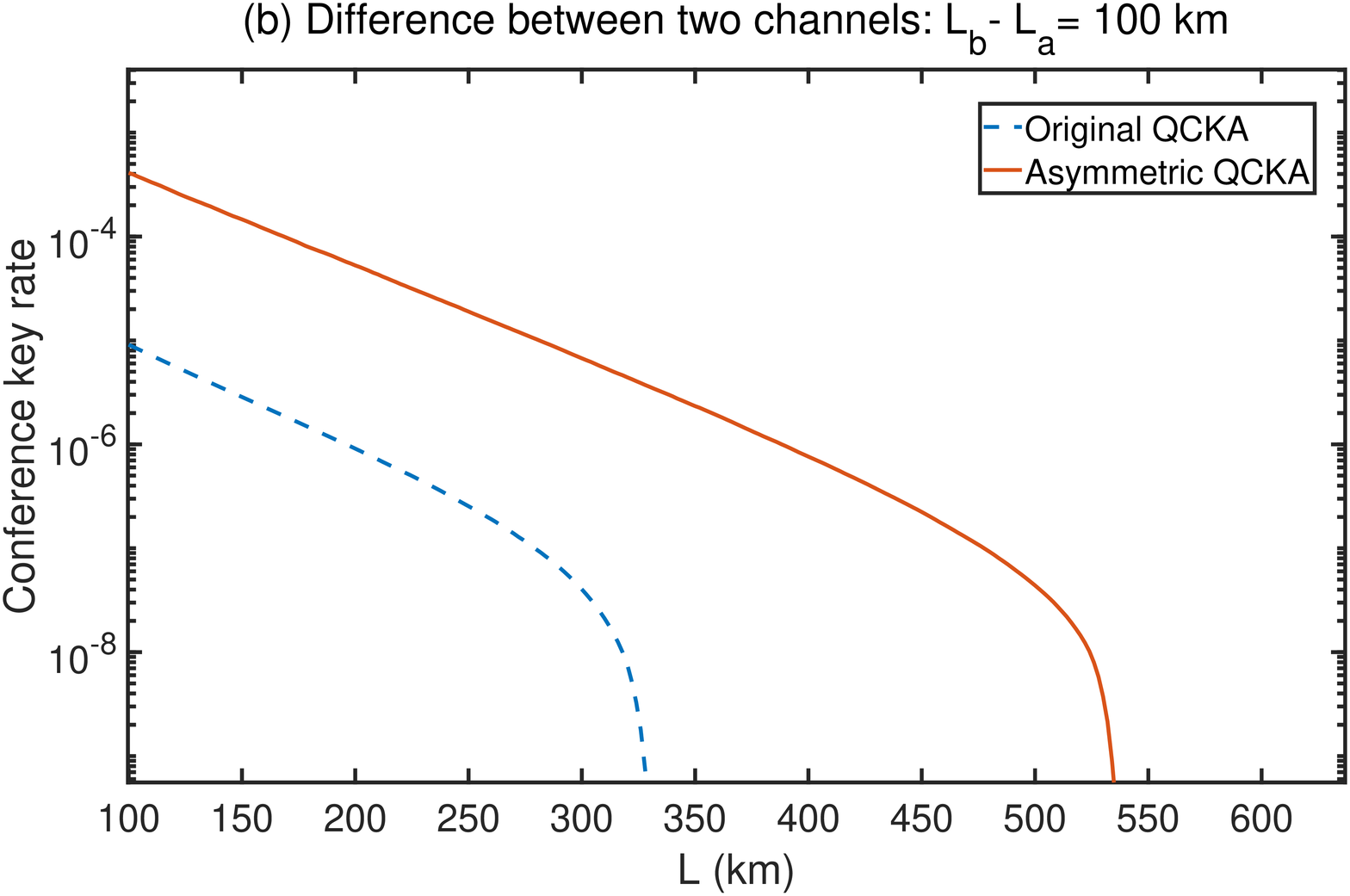}\label{B}
	
	\caption{Comparison of the secure conference key rate between our protocol and the original protocol considered finite-key analysis. The abscissa axis $L=L_a+L_b$ means transmission distance between three parties. (a) Here, the difference in length between two channels is fixed at 50 km. The typical experimental parameters are listed in table \ref{parameters}. Numerically optimized secret key rates with logarithmic scale are obtained for a predetermined signals $N=10^{14}$ with $\varepsilon_{\rm sec}=\varepsilon_{\rm cor}=10^{-10}$. (b) Here, the difference in length between two channels is fixed at 100 km with the same experimental parameters as (a). 
	}
	\label{n14}
\end{figure}

Here, we present the simulation results of different QCKA protocols over asymmetric channels in figure \ref{wxmc} and \ref{n14}. The experimental parameters used in the simulation are listed in table \ref{parameters}. In the asymptotic case, the superiority of our asymmetric protocol is demonstrated in figure \ref{wxmc}. In figure \ref{n14}, the finite-key effect has been considered. 

In the original protocol, the intensities of the pulses interfering at BS3 will differ a lot under different transmission distances, which will result in a relatively high bit error rate in X basis. Therefore, the phase error rate of joint single-photon in Z basis, $\overline{e}_{1}^{\mathrm{ph}}$, will increase a lot. 

In figure \ref{n14}, we show the conference key rates of two protocols with the length difference between the two channels $L_b-L_a$ fixed at a constant length. Here, we assume that the efficiency and dark count rate of Charlie's detectors are the same and define the channel transmittance $\eta$ as $\eta_{d}\times 10^{-\alpha L/10}$. We numerically optimize the conference key rate over the free parameters
$p_{z_a}$, $p_{z_b}$, $\nu_a$, $\nu_b$, $\mu_a$, $\mu_b$, $t_a$, $p_{0_a}$, $p_{0_b}$, $p_{v_a}$, $p_{v_b}$, $q_z$ and $\delta$, where $q_z$ is the beam splitting ratio of BS1 and BS2 for one photon to Z basis. And $t_b$ will be restricted by equation (\ref{mathematical constraint}). Numerically optimized secret key rates are obtained for a predetermined number of signals $N=10^{14}$ with $\varepsilon_{\rm sec}=\varepsilon_{\rm cor}=10^{-10}$. It is easy to find that in the asymmetric channels, our asymmetric protocol outperforms the original one. Moreover, comparing figure \ref{A} with figure \ref{B}, we can notice that the advantage of our asymmetric protocol enhances performance significantly as the difference in distance between two channels increases.

\begin{table}[h]
	\centering  
	\caption{Simulation parameters. $\eta_{d}$ and $p_{d}$ are the detector efficiency and dark count rate. $e_{d}^{x}$ is the misalignment rate of X basis. $\alpha$ is the attenuation coefficient of the ultralow-loss fiber. $f$ is the error correction efficiency. }  
	\begin{tabular}{lcccccc}  
		\hline\hline
		$\eta_{d}$ & $p_{d}$ & $e_{d}^{x}$ & $\alpha$ & $f$ \\
		\hline
		$56\%$ & $10^{-8}$ & $3.5 \%$ & $0.167$dB/km & $1.1$ \\
		\hline\hline
	\end{tabular}  
	\label{parameters}
\end{table} 

\section{Conclusion}
In summary, we propose a QCKA protocol with different source parameters to increase the conference key rate in asymmetric channels. We provide a mathematical constraint in equation (\ref{mathematical constraint}) for the intensity and probability of sending on both sides of Alice and Bob to ensure security in the asymmetric case. Furthermore, we perform finite key analysis for practical scenarios. Note that our protocol is not a measurement-device-independent scheme because Charlie will randomly select the measurement basis. And generalizing our protocol to more than three parties is also a worthwhile work. We hope that this issue will be addressed felicitously in the future. 

Simulation results of two protocols show that our asymmetric protocol has a significantly higher key rate than the original protocol under the condition of asymmetric channels. When the difference in distance between two channels is 100 km, the key rate of the asymmetric protocol is at least one order of magnitude higher than that of the original protocol. The transmission distance is approximately 200 km longer. Meanwhile, our protocol can be directly implemented with current available twin-field QKD devices. This work paves the way for large scale deployment of quantum cryptographic network. 
	
\section*{Acknowledgments}
We gratefully acknowledge support from the National Natural Science Foundation of China (under Grant No. 61801420); the Key-Area Research and Development Program of Guangdong Province (under Grant No. 2020B0303040001); the Fundamental Research Funds for the Central Universities (under Grant No. 020414380182).

\end{document}